\documentclass[11pt]{amsart}
\usepackage{fullpage}
\usepackage{graphicx}
\usepackage{amsmath}
\usepackage{amssymb}
\usepackage{amsbsy}
\newcommand{\mat}[1]{\ensuremath\mathbf{#1}}
\newcommand{\tens}[1]{\ensuremath\boldsymbol{#1}}
\newcommand{\vect}[1]{\ensuremath\boldsymbol{#1}}
\renewcommand{\vec}[1]{\vect{#1}}
\begin{document}

\title[Model-Free Data-Driven Mechanics]{Model-Free Data-Driven Methods in Mechanics:\break
  {\small Material Data Identification 
              and Solvers}}


\author[L. Stainier]{Laurent Stainier}
\address{Research Institute in Civil and Mechanical Engineering 
         (GeM - UMR 6183 CNRS/ECN/UN),
         Ecole Centrale Nantes, F-44321 Nantes, France}
\email{Laurent.Stainier@ec-nantes.fr}
\author[A. Leygue]{Adrien Leygue}
\address{Research Institute in Civil and Mechanical Engineering 
         (GeM - UMR 6183 CNRS/ECN/UN),
         Ecole Centrale Nantes, F-44321 Nantes, France}
\email{Adrien.Leygue@ec-nantes.fr}
\author[M. Ortiz]{Michael Ortiz}
\address{Division of Engineering and Applied Sciences, 
         California Institute of Technology, Pasadena, CA 91125, USA}
\email{ortiz@aero.caltech.edu}


\date{June 2019}

\begin{abstract}
This paper presents an integrated model-free data-driven approach to solid mechanics, allowing to perform numerical simulations on structures on the basis of measures of displacement fields on representative samples, without postulating a specific constitutive model. A material data identification procedure, allowing to infer strain-stress pairs from displacement fields and boundary conditions, is used to build a material database from a set of mutiaxial tests on a non-conventional sample. This database is in turn used by a data-driven solver, based on an algorithm minimizing the distance between manifolds of compatible and balanced mechanical states and the given database, to predict the response of structures of the same material, with arbitrary geometry and boundary conditions. Examples illustrate this modelling cycle and demonstrate how the data-driven identification method allows importance sampling of the material state space, yielding faster convergence of simulation results with increasing database size, when compared to synthetic material databases with regular sampling patterns.

\end{abstract}

\maketitle

\section{Introduction}
\label{intro}

Kirchdoerfer and Ortiz \cite{KirchdoerferOrtiz2016,KirchdoerferOrtiz2017,KirchdoerferOrtiz2018} have recently proposed a new class of problems in static and dynamic elasticity, referred to as {\sl Data-Driven problems}, defined on the space of strain-stress field pairs, or phase space. The problems consist of minimizing the distance between a given material data set and the subspace of compatible strain fields and stress fields in equilibrium. Classical solutions are recovered in the case of elasticity and conditions for convergence of Data-Driven solutions corresponding to sequences of material data sets have been derived by Conti {\sl et al.} \cite{ContiMuellerOrtiz2018}. Data-Driven elasticity effectively reformulates the classical initial-boundary-value problem of elasticity directly from material data, thus bypassing the empirical material modelling step altogether. By eschewing empirical models, material modelling empiricism, modelling error and uncertainty are eliminated entirely and no loss of experimental information is incurred.

Data-driven elasticity relies on the availability of a material data set (referred to as a material database in the following), which may be obtained in different ways. Building this database computationally, for example through micro-macro approaches such as FE$^2$ \cite{Feyel1999,Feyel2003} is expensive and might require efficient model order reduction and high dimensional interpolation techniques. From an experimental point of view on the other hand, it is far from trivial to be able to somehow measure strain and stress over a wide range of deformations. Most identification methods rely on the postulate of a constitutive model, for which optimal parameters are obtained that minimize the distance between experimental measures and numerical results, for example obtained by Finite Element \cite{Rethore2010}. Leygue {\sl et al.} \cite{Leygue_etal2018} recently proposed an alternative approach, completely avoiding the postulate of a specific constitutive model, exploiting the distance-minimization paradigm of Data-Driven  elasticity, and hence coined as Data-Driven Identification.

This paper shows how the two approaches naturally combine to provide a consistent Data-Driven strategy, opening new perspectives on the computational design cycle of structural parts. Examples presented here demonstrate the added value of complex tests on technological samples which, when combined with field measurement techniques (e.g.\ DIC, tomography), provide rich material databases and importance sampling of the material state space.

The paper is organized as follows. We start by recalling the essential elements of the distance minimization paradigm for Data-Driven Computational Mechanics (DDCM), in section \ref{sec:ddcm}. Section \ref{sec:algo} discusses the solver's algorithm, showing how it is connected to traditional linear elasticity solvers, opening the path to interfaces with off-the-shelf computational engines (commercial or other). The paper then proceeds to recall the Data-Driven Identification (DDI) algorithm, in section \ref{sec:ddi}, before presenting examples of the application of the two approaches (DDI+DDCM) combined in section \ref{sec:appl}. Material databases of various sizes are first identified from a biaxial loading test on a plate, fitted with several holes to generate heterogeneous mechanical fields. This material database is then used to solve two different boundary-value problems: a plate with a single hole (different geometry and loading than the sample used for identification) and a L-shaped beam. Results obtained by this DDI+DDCM approach are confronted to those obtained by DDCM with purely synthetic material databases, with uniform sampling of material state space, and to reference solutions obtained by classical FEM using a constitutive model. The paper closes with a discussion of the methodology, results and perspectives.

\section{Material data distance minimisation paradigm}
\label{sec:ddcm}

Let us start by recalling the main lines of the Data-Driven Computational Mechanics (DDCM) approach for elasticity, as recently proposed by \cite{KirchdoerferOrtiz2016,KirchdoerferOrtiz2017}. For simplicity, we consider discrete, or discretized, systems consisting of $N$ nodes and $M$ material points. Such systems typically arise from a Finite Element (FE) discretization, but other methods can be included in the framework as well. The system undergoes displacements $\vec{u} = \{\vec{u}_a\}_{a=1}^N$, with $\vec{u}_a \in \mathbb{R}^{n_a}$ and $n_a$ the dimension of the displacement at node $a$, under the action of applied forces $\vec{f} = \{\vec{f}_a\}_{a=1}^N$, with $\vec{f}_a \in \mathbb{R}^{n_a}$. The local state of the system is characterized by stress and strain pairs $\{(\tens{\epsilon}_e, \tens{\sigma}_e)\}_{e=1}^M$, with $\tens{\epsilon}_e, \tens{\sigma}_e \in \mathbb{R}^{m_e}$ and $m_e$ the dimension of stress and strain at material point $e$. We regard $\vec{z}_e = (\tens{\epsilon}_e, \tens{\sigma}_e)$ as a point in a local phase space $Z_e = \mathbb{R}^{m_e} \times \mathbb{R}^{m_e}$ and $\vec{z} = \{(\tens{\epsilon}_e, \tens{\sigma}_e)\}_{e=1}^M$ as a point in the global phase space $Z = Z_1 \times \cdots \times Z_M$.

The internal state of the system is subject to the compatibility and equilibrium constraints of the general form
\begin{subequations}\label{eq:constraints}
\begin{align}
    & \label{eq:compat}
    \tens{\epsilon}_e = \mat{B}_e \vec{u} ,
    \quad e=1,\dots,M,
    \\ & \label{eq:equil}
    \sum_{e=1}^M w_e \mat{B}_e^T \tens{\sigma}_e = \vec{f} ,
\end{align}
\end{subequations}
where $\{w_e\}_{e=1}^M$ are elements of volume and $\mat{B}_e$ is a discrete strain operator for material point $e$. We note that constraints \eqref{eq:constraints} are universal, or material-independent. They define a subspace, or constraint set,
\begin{equation}\label{eq:equil_set}
    E = \{ \vec{z} \in Z \, : \, (\ref{eq:compat}) \text{ and } (\ref{eq:equil}) \} ,
\end{equation}
consisting of all compatible and equilibrated internal states. In \eqref{eq:equil_set} and subsequently, the symbol $:$ is used to mean 'given' or 'subject to' or 'conditioned to'. Within this subspace, the local state satisfies the power identity
\begin{equation}
    \vec{f} \cdot \vec{u}
    =
    \sum_{e=1}^M w_e \, \tens{\sigma}_e \cdot \tens{\epsilon}_e .
\end{equation}

In classical elasticity, the problem \eqref{eq:constraints} is closed by appending local material laws, e.~g., functions of the general form
\begin{equation}\label{eq:constitutive}
    \tens{\sigma}_e = \hat{\tens{\sigma}}_e(\tens{\epsilon}_e) ,
    \quad
    e = 1,\dots, M ,
\end{equation}
where $\hat{\tens{\sigma}}_e : \mathbb{R}^{m_e} \to \mathbb{R}^{m_e}$. However, often material behavior is only known through a material data set $D_e$ of points $\vec{z}_e = (\tens{\epsilon}_e,\tens{\sigma}_e) \in Z_e$ obtained experimentally or by some other means. Again, the conventional response to this situation is to deduce a material law $\hat{\tens{\sigma}}_e$ from the data set $D_e$ by some appropriate means, thus reverting to the classical setting \eqref{eq:constitutive}.

The Data-Driven reformulation of the classical problems of mechanics consists of formulating boundary-value problems directly in terms of the material data, thus entirely bypassing the material modeling step altogether \cite{KirchdoerferOrtiz2016}. A class of Data-Driven problems consists of finding the compatible and equilibrated internal state $\vec{z} \in E$ that minimizes the distance to the global material data set $D = D_1 \times \cdots \times D_M$. To this end, we metrize the local phase spaces $Z_e$ by means of norms of the form
\begin{equation}\label{eq:local_norm}
    | \vec{z}_e |_e
    =
    \left(
        \mathbb{C}_e \tens{\epsilon}_e \cdot \tens{\epsilon}_e
        +
        \mathbb{C}_e^{-1} \tens{\sigma}_e \cdot \tens{\sigma}_e
    \right)^{1/2} ,
\end{equation}
for some symmetric and positive-definite matrices $\{\mathbb{C}_e\}_{e=1}^M$, with corresponding distance
\begin{equation}
    d_e(\vec{z}_e,\vec{y}_e) = | \vec{z}_e - \vec{y}_e |_e ,
\end{equation}
for $\vec{y}_e, \vec{z}_e \in Z_e$. The local norms induce a metrization of the global phase $Z$ by means of the global norm
\begin{equation}\label{eq:global_norm}
    | \vec{z} |
    =
    \Big( \sum_{e=1}^M w_e | \vec{z}_e |_e^2 \Big)^{1/2} ,
\end{equation}
with associated distance
\begin{equation}
    d(\vec{z},\vec{y}) = |\vec{z} - \vec{y}|,
\end{equation}
for $\vec{y}, \vec{z} \in Z$. The distance-minimizing Data-Driven problem is, then,
\begin{equation}\label{eq:dist_min}
    \min_{\vec{y} \in D} \; \min_{\vec{z} \in E} \; d(\vec{z},\vec{y})
    =
    \min_{\vec{z} \in E} \; \min_{\vec{y} \in D} \; d(\vec{z},\vec{y}) ,
\end{equation}
i.~e., we wish to find the point $\vec{y} \in D$ in the material data set that is closest to the constraint set $E$ of compatible and equilibrated internal states or, equivalently, we wish to find the compatible and equilibrated internal state $\vec{z} \in E$ that is closest to the material data set $D$.

We emphasize that the local material data sets can be graphs, point sets, 'fat sets' and ranges, or any other arbitrary set in phase space. Evidently, the classical problem is recovered if the local material data sets are chosen as
\begin{equation}
    D_e
    =
    \{
        (\tens{\epsilon}_e,\hat{\tens{\sigma}}_e(\tens{\epsilon}_e))
    \} ,
\end{equation}
i.~e., as \emph{graphs} in $Z_e$ defined by the material law \eqref{eq:constitutive}. Thus, the Data-Driven reformulation \eqref{eq:dist_min} extends --and subsumes as special cases-- the classical problems of mechanics.

\section{Data-Driven simulation algorithm}
\label{sec:algo}

Note that, for fixed $\vec{y} \in D$, the closest point projection $\vec{z} = P_E \vec{y}$ onto $E$ follows by minimizing the quadratic function $d^2(\cdot,\vec{y})$ subject to the constraints \eqref{eq:constraints}. The compatibility constraint \eqref{eq:compat} can be enforced directly by introducing a displacement field $\vec{u}$. The equilibrium constraint \eqref{eq:equil} can then be enforced by means of Lagrange multipliers $\vec{\lambda}$ representing virtual displacements of the system. With $\vec{y} \equiv \{(\tens{\epsilon}^*_e, \tens{\sigma}^*_e)\}_{e=1}^M$ given, e.~g.\ from a previous iteration, the corresponding Euler-Lagrange equations are \cite{KirchdoerferOrtiz2016}
\begin{subequations}
\label{eq:DDCM-equations}
\begin{align}
  \label{eq:DDCM-u}
    &
    \Big(\sum_{e=1}^M w_e \mat{B}_e^T \mathbb{C}_e \mat{B}_e \Big) \vec{u}
    =
    \sum_{e=1}^M w_e \mat{B}_e^T \mathbb{C}_e \tens{\epsilon}^*_e ,
    \\ \label{eq:DDCM-lambda} &
    \Big(\sum_{e=1}^M w_e \mat{B}_e^T \mathbb{C}_e \mat{B}_e \Big) \vec{\lambda}
    =
    \vec{f} - \sum_{e=1}^M w_e \mat{B}_e^T \tens{\sigma}^*_e ,
\end{align}
\end{subequations}
which define two standard linear displacement problems. The closest point $\vec{z} = P_E \vec{y} \in E$ then follows as
\begin{subequations}
\begin{align}
    \label{eq:DDCM-eps} &
    \tens{\epsilon}_e = \mat{B}_e \vec{u} ,
    \quad
    e = 1,\dots,M ,
    \\ \label{eq:DDCM-sigma} &
    \tens{\sigma}_e = \tens{\sigma}^*_e + \mathbb{C}_e \mat{B}_e \vec{\lambda} ,
    \quad
    e = 1,\dots,M .
\end{align}
\end{subequations}
A simple Data-Driven solver then consists of the fixed point iteration \cite{KirchdoerferOrtiz2016}
\begin{equation}\label{eq:DD_solver}
    \vec{z}_{j+1} = P_E P_D \vec{z}_j ,
\end{equation}
for $j = 0,1,\dots$ and $\vec{z}_0 \in Z$ arbitrary, where $P_D$ denotes the closest point projection of a point in $Z$ onto $D$. Iteration \eqref{eq:DD_solver} first finds the closest point $P_D \vec{z}_j$ to $\vec{z}_j$ on the material data set $D$ and then projects the result back to the constraint set $E$. The iteration is repeated until $P_D \vec{z}_{j+1} = P_D \vec{z}_j$, i.~e., until the data association to points in the material data set remains unchanged. Note that more sophisticated algorithms can be considered to solve this combinatorial optimization problem \cite{Kanno2019}.

Equations \eqref{eq:DDCM-equations} define two standard linear elasticity problems, which can be interpreted as follows. Equation \eqref{eq:DDCM-u} states that displacement field should be compatible with material strains $\{\tens{\epsilon}^*_i\}$ in a weak sense, i.~e.\ strains computed from $\vec{u}$ at material points associated to the same data point $\tens{\epsilon}^*_i$ should average to that value. In view of \eqref{eq:DDCM-lambda} or \eqref{eq:DDCM-sigma}, Lagrange multipliers $\vec{\lambda}$ can be interpreted as a discrete displacement field corresponding to the mismatch between mechanical and material stresses, resp.\ $\tens{\sigma}_e$ and $\tens{\sigma}^*_e$. From a practical point of view, the above linear elasticity problems can be treated as classical problems, with arbitrary elastic properties $\mathbb{C}_e$ (possibly non-homogeneous), and subject to eigen-strain or eigen-stress fields. In particular, system \eqref{eq:DDCM-u} corresponds to a linear elasticity problems where the physical kinematic boundary conditions are enforced, and where the only other loading consists in the eigen-strain field described by $\{\tens{\epsilon}^*_e\}$, i.~e.\ where applied body forces and (non-zero) static boundary conditions of the physical problem are not enforced. On the other hand, system \eqref{eq:DDCM-lambda} corresponds to a linear elasticity problem with homogeneous kinematic boundary conditions, where the physical loading is applied (body forces and static boundary conditions) together with a field or eigen-stresses (or residual stresses) described by $\{\tens{\sigma}^*_e\}$. Ideally, this eigen-stress field should be balanced with the applied loading, yielding null Lagrange multipliers. Finally note that the linearity of systems \eqref{eq:DDCM-equations} to be solved is completely independent of the linearity or non-linearity of the material behavior described by $D$.

The convergence properties of the fixed-point solver \eqref{eq:DD_solver} have been investigated in \cite{KirchdoerferOrtiz2016}. The Data-Driven paradigm has been extended to dynamics \cite{KirchdoerferOrtiz2018}, finite kinematics \cite{NguyenKeip2018} and objective functions other than phase-space distance can be found in \cite{KirchdoerferOrtiz2017}. The well-posedness of Data-Driven problems and properties of convergence with respect to the data set have been investigated in \cite{ContiMuellerOrtiz2018}.

\section{Material data identification}
\label{sec:ddi}

The DDCM paradigm exposed in section \ref{sec:ddcm} relies critically on the availability of a material data set $D$. For three-dimen\-sional elasticity, sufficient phase-space coverage (i.~e. importance sampling) may require a very large number of data points, which may not be easily amenable by classical mechanical testing (uniaxial, biaxial or shear loadings). To address this challenge, a material data set can be directly constructed, as proposed by \cite{Leygue_etal2018}, from a collection of displacement and (non homogeneous) strain fields, associated with a series of known boundary conditions, i.~e.\ imposed displacements and/or (resultant) forces. These fields could for example be obtained by Digital Image Correlation (DIC). This Data-Driven Identification (DDI) method, summarized below, simultaneously identifies the stress component of the mechanical state for each loading condition and the full material states database, which is common to all loading conditions.

For each data item $\alpha$ (or snapshot), the following quantities are available:
\begin{itemize}
\item discrete (e.~g.\ nodal) displacements $\vec{u}^\alpha = \{\vec{u}^\alpha_a\}_{a=1}^N$,
\item the (discretized) geometry, encoded through matrices
$\mat{B}^\alpha_e$, which can compute the mechanical strain $\tens{\epsilon}^\alpha_e = \mat{B}^\alpha_e \vec{u}^\alpha$,
\item applied nodal forces $\vec{f}^\alpha$,
\item prescribed nodal displacements.
\end{itemize}
The aim of the DDI technique is to compute a number $N^*$ of material states $(\tens{\epsilon}^*_i , \tens{\sigma}^*_i )$ such that:
\begin{itemize}
\item for each snapshot $\alpha$ and material point $e$, it is possible to compute the mechanical state $\tens{\sigma}^\alpha_e$ which satisfies mechanical equilibrium,
\item for each snapshot, a material state $(\tens{\epsilon}^*_{ie},\tens{\sigma}^*_{e})$ can be assigned to each material point $e$ which is the closest to the mechanical state according to norm \eqref{eq:local_norm}.
\end{itemize}
It results the following minimization problem
\begin{equation}
 \label{eq:DDI}
  \min_{\tens{\sigma}^\alpha_e,\tens{\epsilon}^*_i,\tens{\sigma}^*_i, ie^\alpha} 
    \sum_\alpha d(\{(\tens{\epsilon}^\alpha_e, \tens{\sigma}^\alpha_e)\},
                            \{(\tens{\epsilon}^*_{ie^\alpha}, \tens{\sigma}^*_{ie^\alpha})\}) ,
\end{equation}
subject to constraints
\begin{equation}
  \label{eq:equil-DDI}
  \sum_{e=1}^M w_e^\alpha {\mat{B}^\alpha_e}^T \tens{\sigma}^\alpha_e = \vec{f}^\alpha 
  \quad\forall \alpha .
\end{equation}
In \eqref{eq:DDI}, $ie^\alpha$ denotes a mapping between material points ($e$) and data points ($i$) for snapshot $\alpha$. For an arbitrary state mapping $ie^\alpha$, the equilibrium constraints \eqref{eq:equil-DDI} are enforced by means of Lagrange multipliers $\vec{\eta}^\alpha$, yielding the following stationarity equations \cite{Leygue_etal2018}:
\begin{subequations}
\label{eq:DDI-system}
\begin{align}
    &\Big(\sum_{e=1}^M w^\alpha_e {\mat{B}^\alpha_e}^T \mathbb{C}_e \mat{B}^\alpha_e \Big) \vec{\eta}^\alpha
    - \sum_{e=1}^M w^\alpha_e {\mat{B}^\alpha_e}^T \tens{\sigma}^*_{ie^\alpha(e)}
    =
    \vec{f}^\alpha \quad \forall\alpha , \\
    &\sum_\alpha \sum_{e:ie^\alpha(e)=i} w^\alpha_e \mat{B}^\alpha_e \vec{\eta}^\alpha = 0
    \quad\forall i \in [1:N^*] .
\end{align}
\end{subequations}
This linear system of equations is solved to simultaneously determine $\{\tens{\sigma}^*_i\}_{i=1}^{N^*}$ and $\{\tens{\sigma}^\alpha_e\}_{e=1}^M\;(\forall\alpha)$ (through $\vec{\eta}^\alpha$, see \cite{Leygue_etal2018}). 
The following fixed-point iterations algorithm is then used to compute the material data set, mechanical stresses, and the state mapping:
\begin{enumerate}
	\item simultaneously initialize $\{\tens{\epsilon}_i^*\}$ and $\{ie^\alpha\}$ by a \emph{k-means} algorithm on $\{\tens{\epsilon}_e^\alpha\}$,
	\item simultaneously compute $\{\tens{\sigma}_i^*\}$ and $\{\vec{\eta}^\alpha\}$ from \eqref{eq:DDI-system},
	\item update the value of $\{\tens{\sigma}_e^\alpha\}$ as
	\begin{equation}
	  \tens{\sigma}_e^\alpha = \tens{\sigma}^*_{ie^\alpha} + \mat{B}^\alpha_{ie^\alpha}\vec{\eta}^\alpha ,
	\end{equation}
	\item compute new state mappings $\{ie^\alpha\}$ with:
	\begin{equation}\label{eq:compute_ie2}
	  \{ie^\alpha\} = \arg \min_{ie^\alpha} 
	                        \sum_\alpha d(\{(\tens{\epsilon}^\alpha_e, \tens{\sigma}^\alpha_e)\},
                                                    \{(\tens{\epsilon}^*_{ie^\alpha}, \tens{\sigma}^*_{ie^\alpha})\}) ,
	\end{equation}
	\item update $\{\tens{\epsilon}_i^*\}$ from
	\begin{equation}
	  \sum_\alpha \sum_{e:ie^\alpha(e)=i} w_e^\alpha 
	            \mathbb{C}_e (\tens{\epsilon}^\alpha_e - \tens{\epsilon}^*_{ie^\alpha} ) = 0 ,
	\end{equation}
	\item iterate steps 2--5 until convergence of $\{ie^\alpha\}$.
\end{enumerate}
Note that the most numerically expensive part are steps 2 and 4, which involve the solution of a large linear system and a database search, respectively.

\section{Application examples}
\label{sec:appl}

In the following, we present a complete cycle of material data identification from strain fields obtained from a set of mechanical tests on a representative sample, followed by data-driven simulation of a different mechanical problem (involving the same material, of course) using the data identified in the first step.

\subsection{Data identification}

\begin{figure}[htb]
\centering
\includegraphics[width=0.5\linewidth]{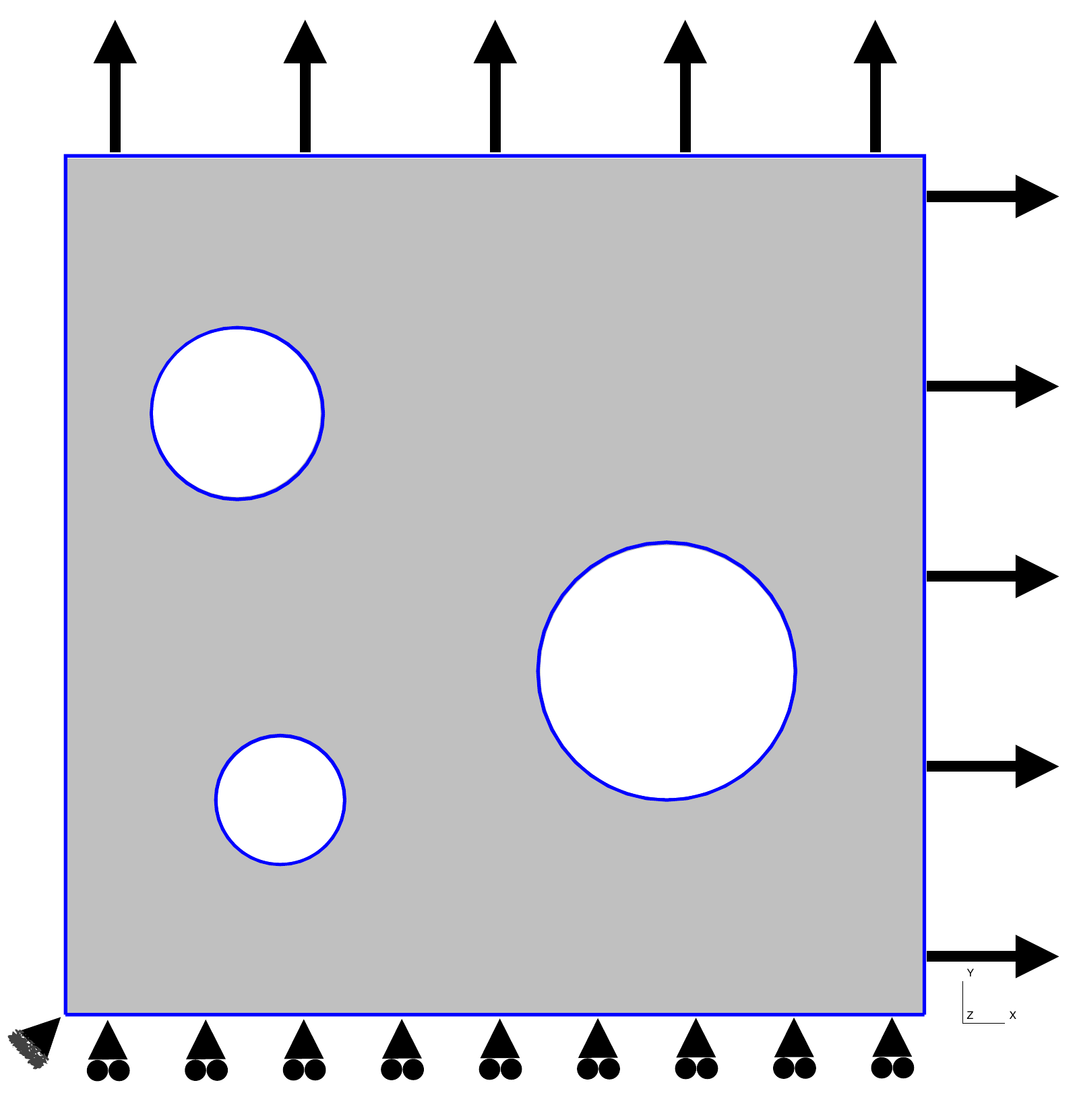}
\caption{Biaxial test sample used for Data Identification}
\label{fig:sample1}       
\end{figure}

For data identification, we use displacement fields and reaction force curves obtained from (displacement controlled) tests on the sample illustrated in Fig.~\ref{fig:sample1}. In the present case, these tests were performed numerically, on a linear isotropic elastic material ($E=217.5$ GPa, $\nu=0.3$, plane strain), but such displacement fields could come from DIC measures as well. Numerically generated results will be free of experimental noise, but in the following we will focus more on questions of importance sampling than on the effect of noisy data.
 
\begin{figure}[htb]
\includegraphics[width=0.5\linewidth]{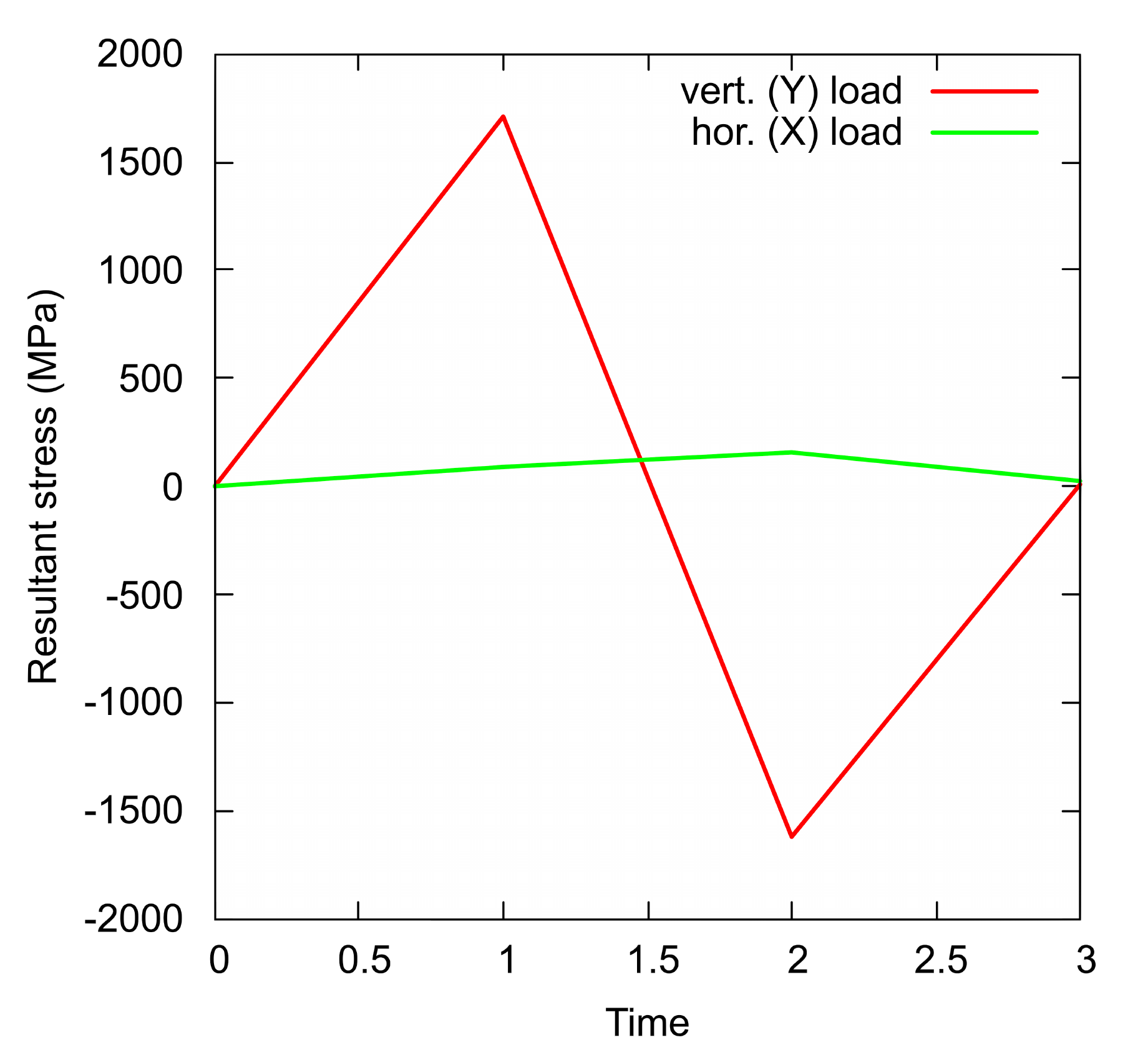}
\caption{Biaxial loading of sample used for Data Identification}
\label{fig:sample1-load}       
\end{figure}

\begin{figure*}
\centering
\includegraphics[width=0.4\textwidth]{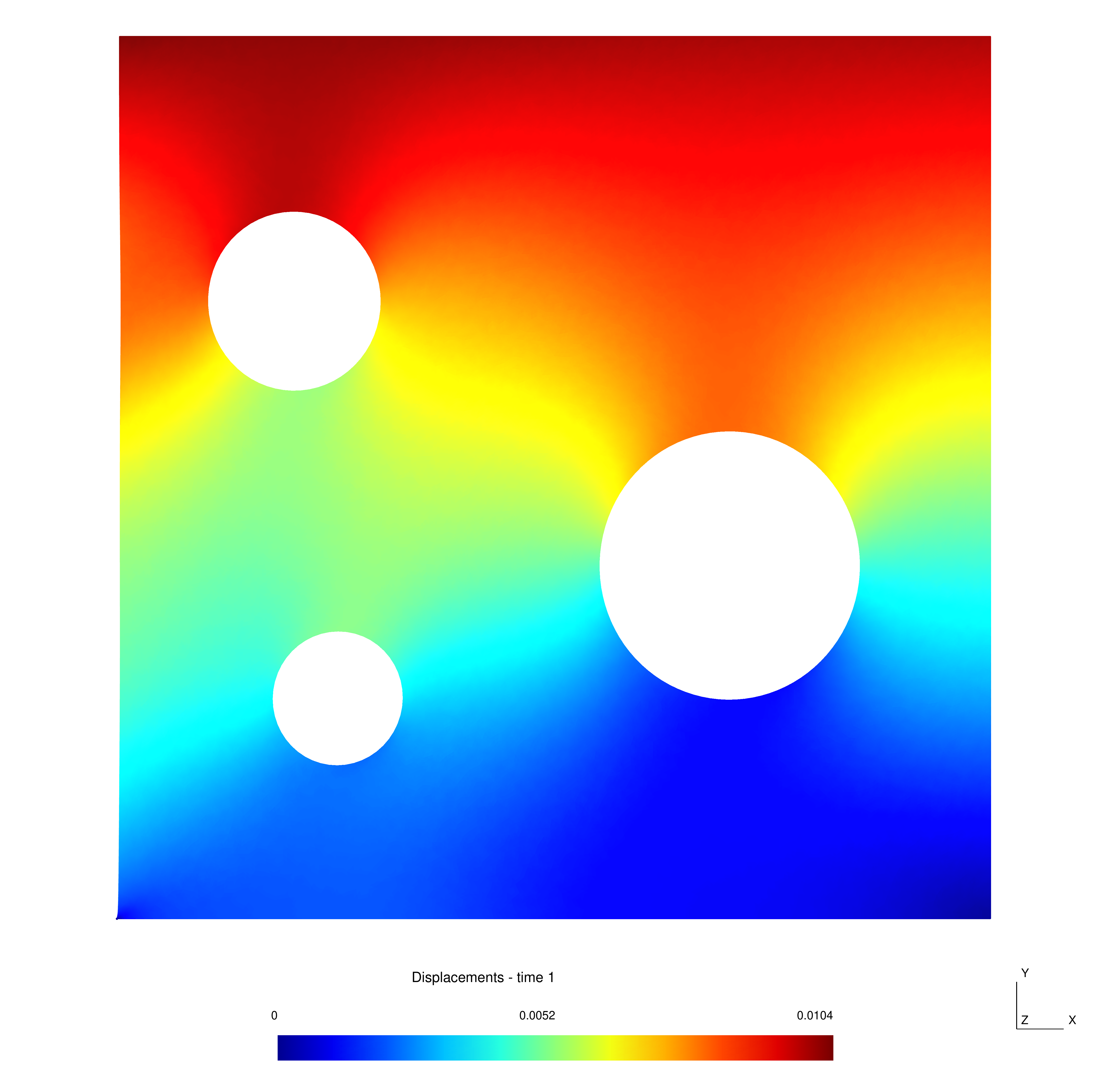}
\hfil
\includegraphics[width=0.4\textwidth]{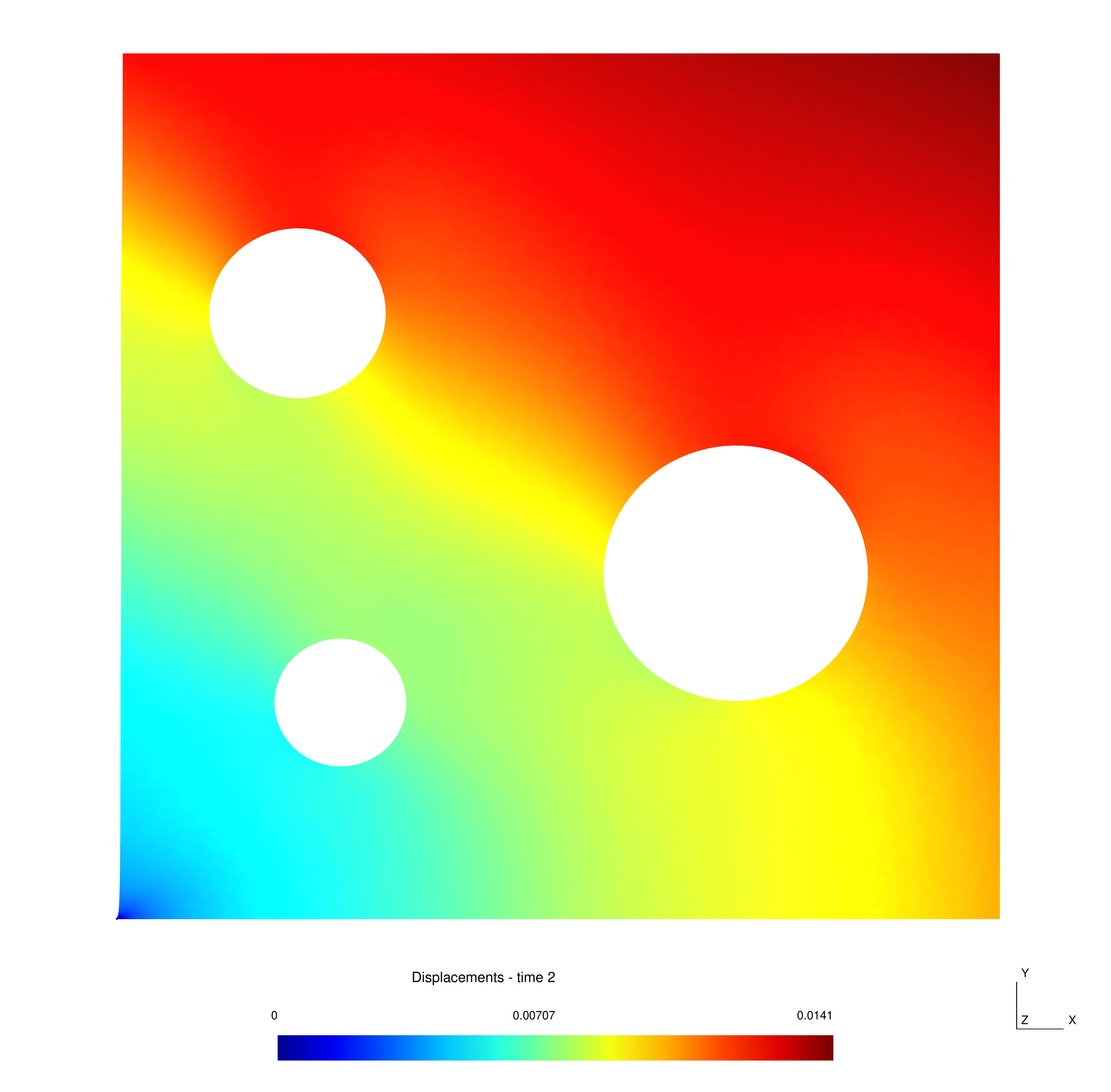}
\caption{Representative snapshots of displacement fields used for Data Identification}
\label{fig:sample1-snapshots}       
\end{figure*}

Figure \ref{fig:sample1-load} provides resultant loads in both directions, as a function of time. This information will be used, together with snapshots of displacement fields, to construct the data\-base. Two examples of displacement snapshots (at times $t=1$ and $t=2$, corresponding to peak loads) are illustrated at Fig.~\ref{fig:sample1-snapshots}.

\begin{figure}[htb]
\includegraphics[width=0.9\linewidth]{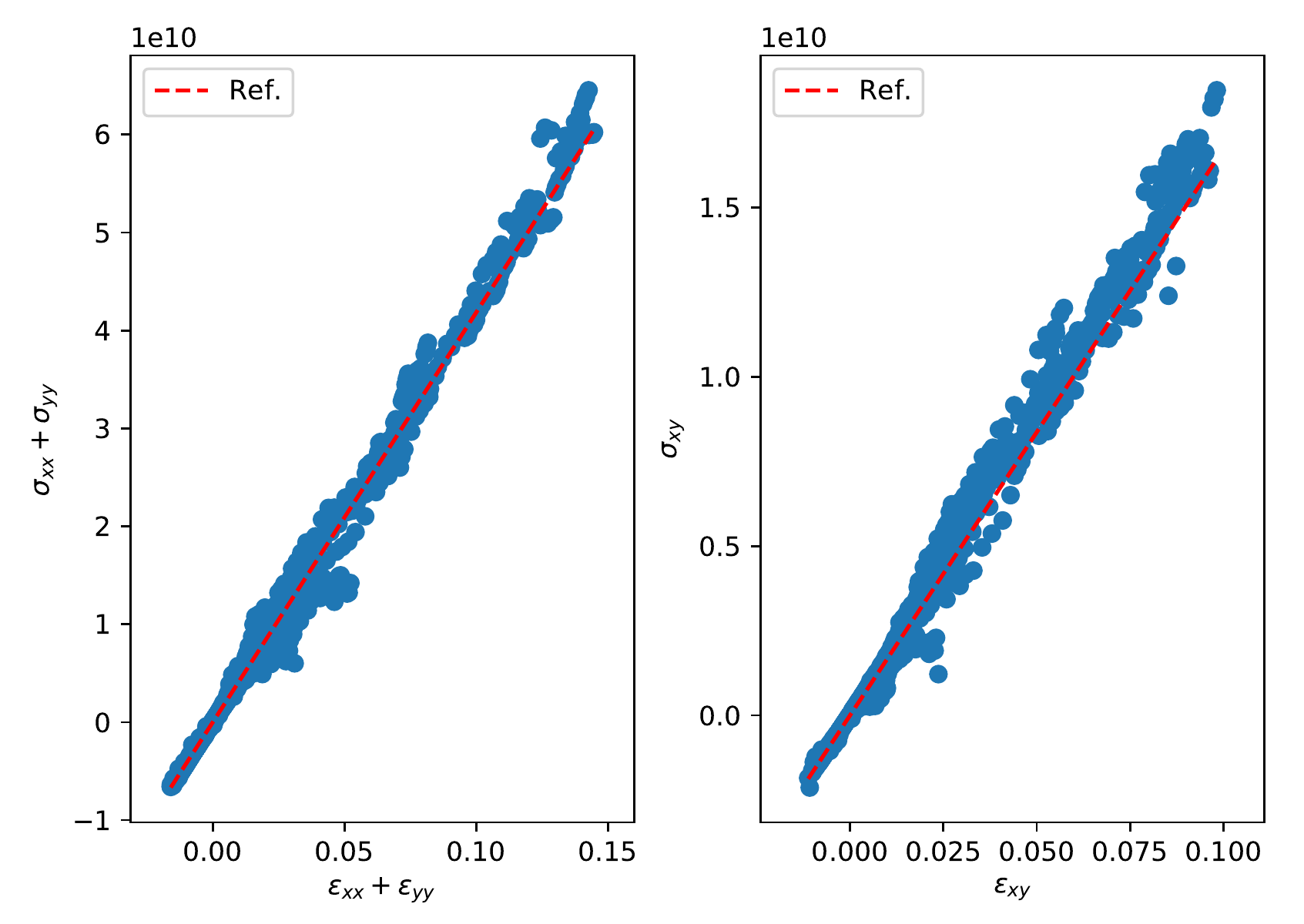}
\caption{Set of srain-stress states in material database DDI-DB2}
\label{fig:DB2}       
\end{figure}

Three different databases have been constructed using the DDI algorithm. They will be referred to as DDI-DB1, DDI-DB2, and DDI-DB3, and contain 10\,000, 25\,000, and 100\,000 data points (i.e.\ strain-stress pairs), respectively. The typical ratio in number of measured mechanical states ($M$ times number of load cases) to number of identified material states ($N^*$) which ensures good performance of the DDI algorithm is about 200. The number of measured mechanical states can be increased by considering more snapshots and/or higher resolution displacement fields, in order to be able to evaluate gradients at more points in space. Figure \ref{fig:DB2} shows the set of strain-stress pairs contained in DDI-DB2, with a comparison to the reference linear elastic behavior. We see that the identification process leads to significant variations around this reference behavior, the level of this ``noise'' being related to the heterogeneity of the input fields.
 
\begin{figure}[htb]
\includegraphics[width=0.7\linewidth]{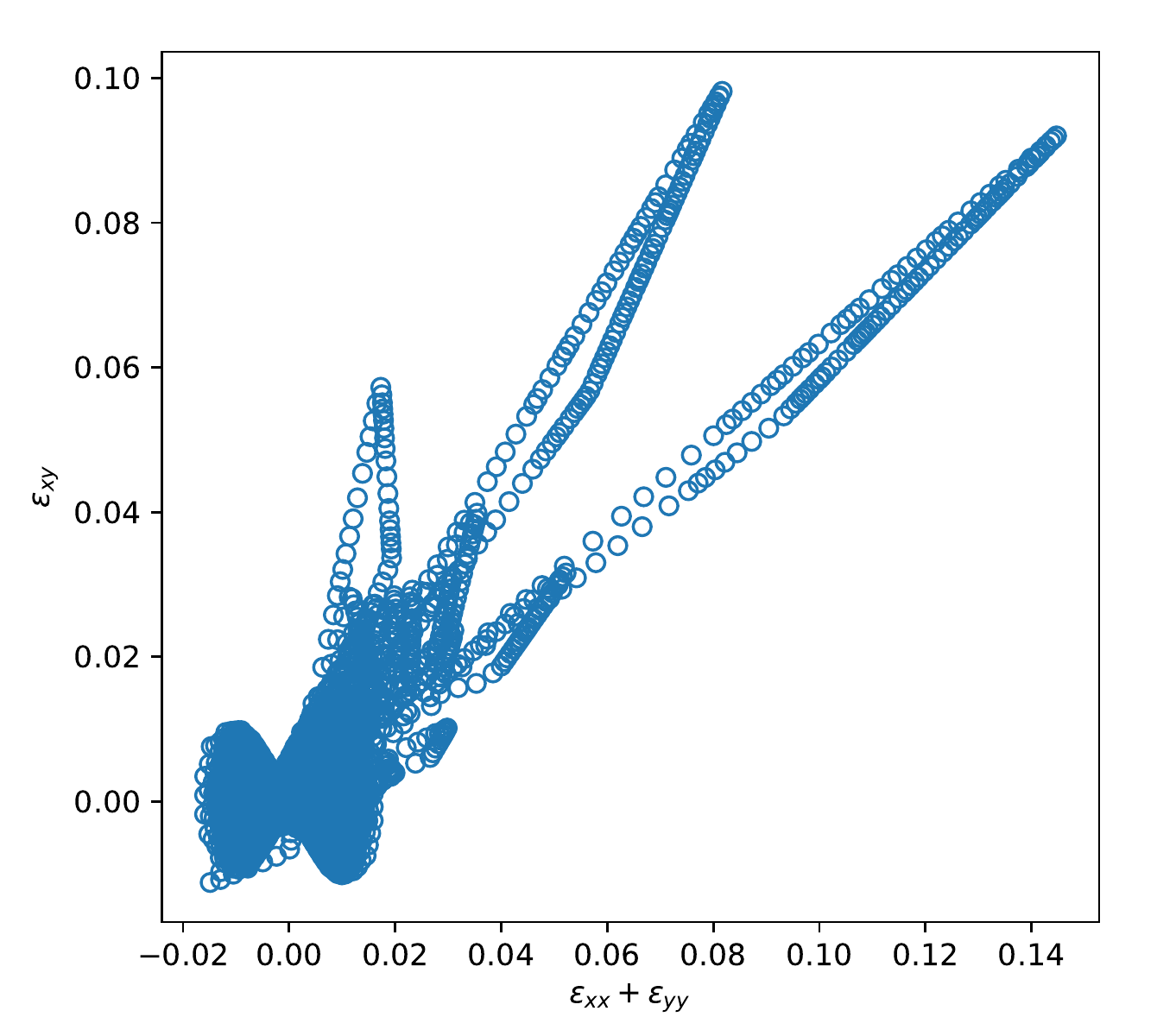}
\caption{Sampling of strain space for material database DDI-DB2}
\label{fig:DB2-strain-sampling}       
\end{figure}

In the DDI framework, the sampling of strain space is dictated by the mechanics of the chosen test. As a way of illustration, the set of strains corresponding to database DDI-DB2 is shown in Fig.~\ref{fig:DB2-strain-sampling}. One can observe that the loading is not purely radial, and also that the density of points is not uniform. The bulk of material points are concentrated around moderate values, with a few points, corresponding to the strain concentration in the lower left corner, exploring larger strain values. 
 
\subsection{Data-driven simulation}

All following DDCM results were obtained with a uniform metric $\mathbb{C}_e$, corresponding to an isotropic Hookean material with $E=100$ GPa and $\nu=0.35$.

\subsubsection{Plate with a hole}

As a first example, we consider a thick plate (plane strain assumption) of width 12.8 $R$ and height 20 $R$, where $R$ is the radius of a circular hole located in the middle of the plate. By symmetry, only one quarter of the plate is modelled. The plate is subjected to a compression load in the vertical direction (average longitudinal strain of -0.4\%). A reference solution is obtained by using a classical FE simulation, using the same elastic properties as those used to generate the input images for the DDI method above (i.e.\ $E=217.5$ GPa, $\nu=0.3$, plane strain), on a mesh of 492 bilinear quadrangular elements (450 nodes). The total elastic strain energy associated to the reference FE solution is 737 J. This number will constitute a useful reference when discussing the distances to and from material databases in the following.

\begin{figure}[htb]
\centering
\includegraphics[width=\linewidth]{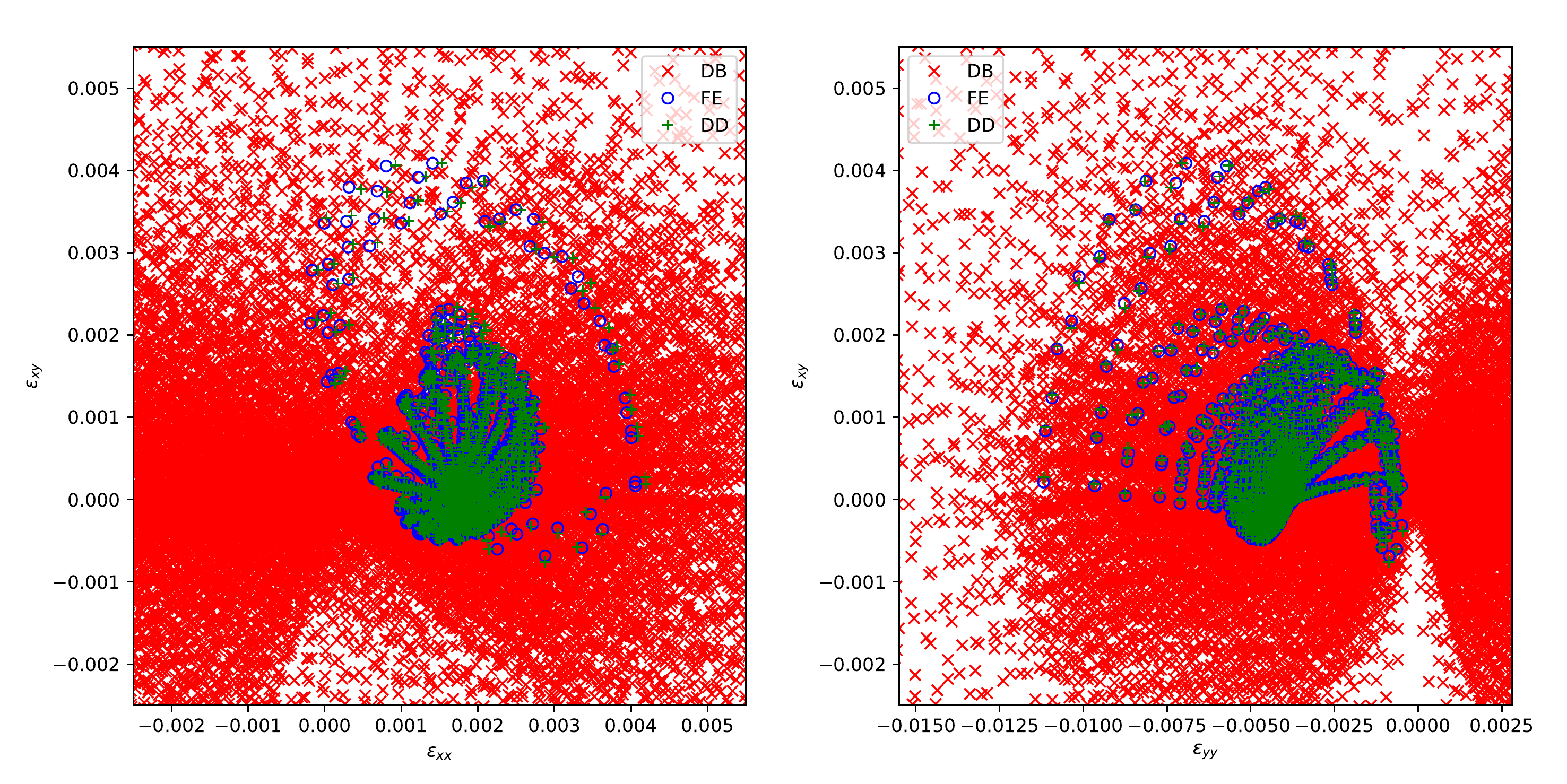}
\caption{Sampling of strain space for material database DDI-DB2, associated DDCM, and FEM solutions}
\label{fig:DDvsFEvsDB-ddi-plate}       
\end{figure}

\begin{figure}[hbt]
\centering
\includegraphics[width=\linewidth]{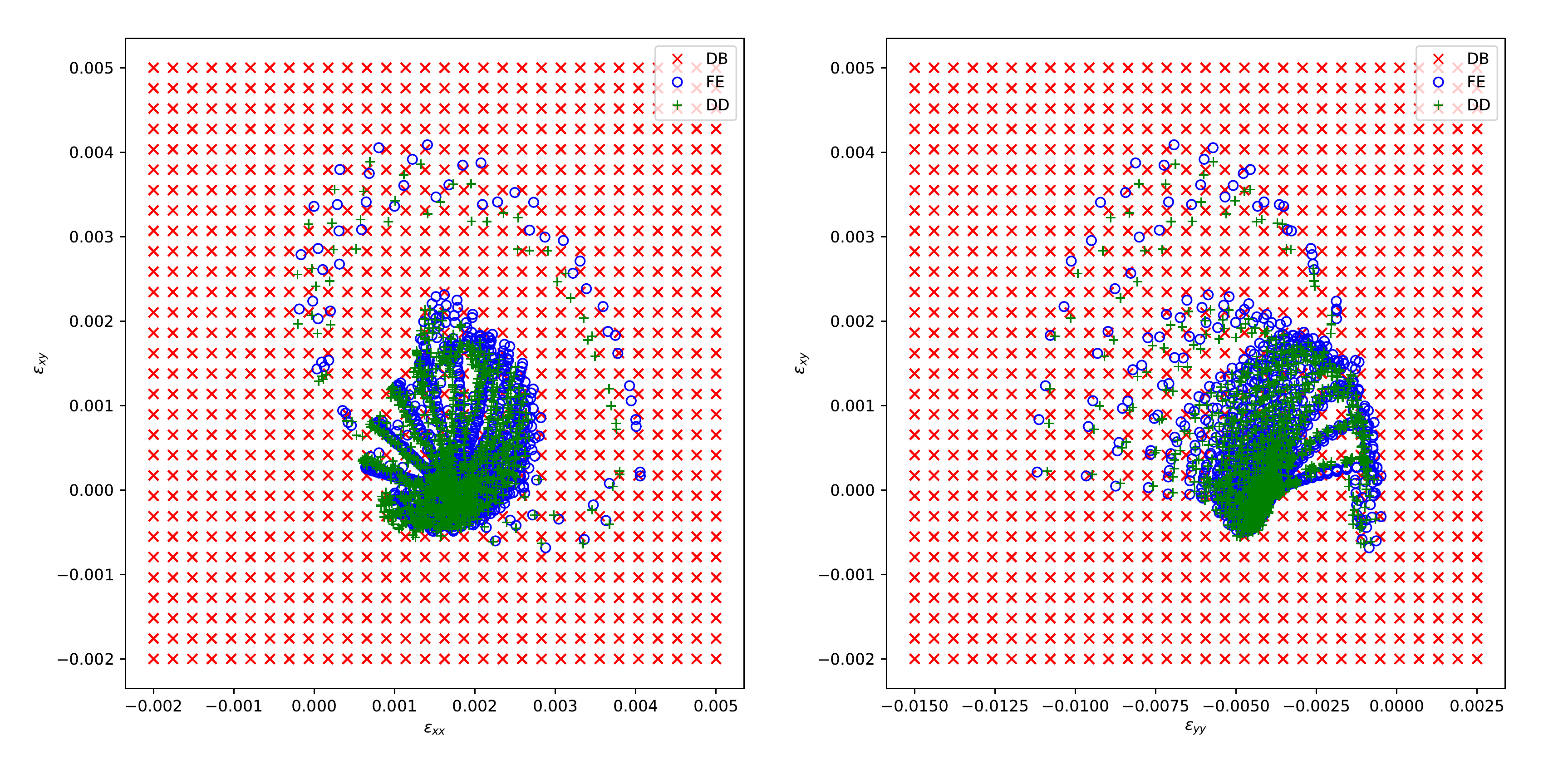}
\caption{Sampling of strain space for material database REG-DB1, associated DDCM, and FEM solutions}
\label{fig:DDvsFEvsDB-reg-plate}       
\end{figure}

This problem was simulated using the DDCM method described in section \ref{sec:algo}, using the different material databases identified by DDI (see above), and the same mesh as for the FE reference solution. For comparison, we also used three purely synthetic databases, labelled REG-DB1, REG-DB2, and REG-DB3, obtained by applying the constitutive equations to uniformly sampled strain tensors: we considered samplings of the $\{\epsilon_{xx},\epsilon_{xy},\epsilon_{yy}\}$ space of $30 \times 30 \times 30$,  $50 \times 50 \times 50$, and  $100 \times 100 \times 100$ points, respectively, in the range specified in Table \ref{tab:eps_range}. Note that these ranges were chosen in function of the strain distribution computed in the FE solution. On the contrary, the range of strains in the databases generated by DDI are not directly controlled.
For reference, Figures \ref{fig:DDvsFEvsDB-ddi-plate} and \ref{fig:DDvsFEvsDB-reg-plate} show how the strain space is spanned by the databases (DDI-DB2 and REG-DB1), FEM and DDCM solutions.

\begin{table}[htb]
\caption{Range of strains in synthetic material databases}
\label{tab:eps_range}
\begin{tabular}{lccc}
\hline\noalign{\smallskip}
Bounds & $\epsilon_{xx}$ & $\epsilon_{xy}$ & $\epsilon_{yy}$ \\
\noalign{\smallskip}\hline\noalign{\smallskip}
min. &-0.002 & -0.002 & -0.015 \\
max. & 0.005 & 0.005 & 0.0025 \\
\noalign{\smallskip}\hline
\end{tabular}
\end{table}


\begin{figure}[htb]
\centering
\includegraphics[width=\linewidth]{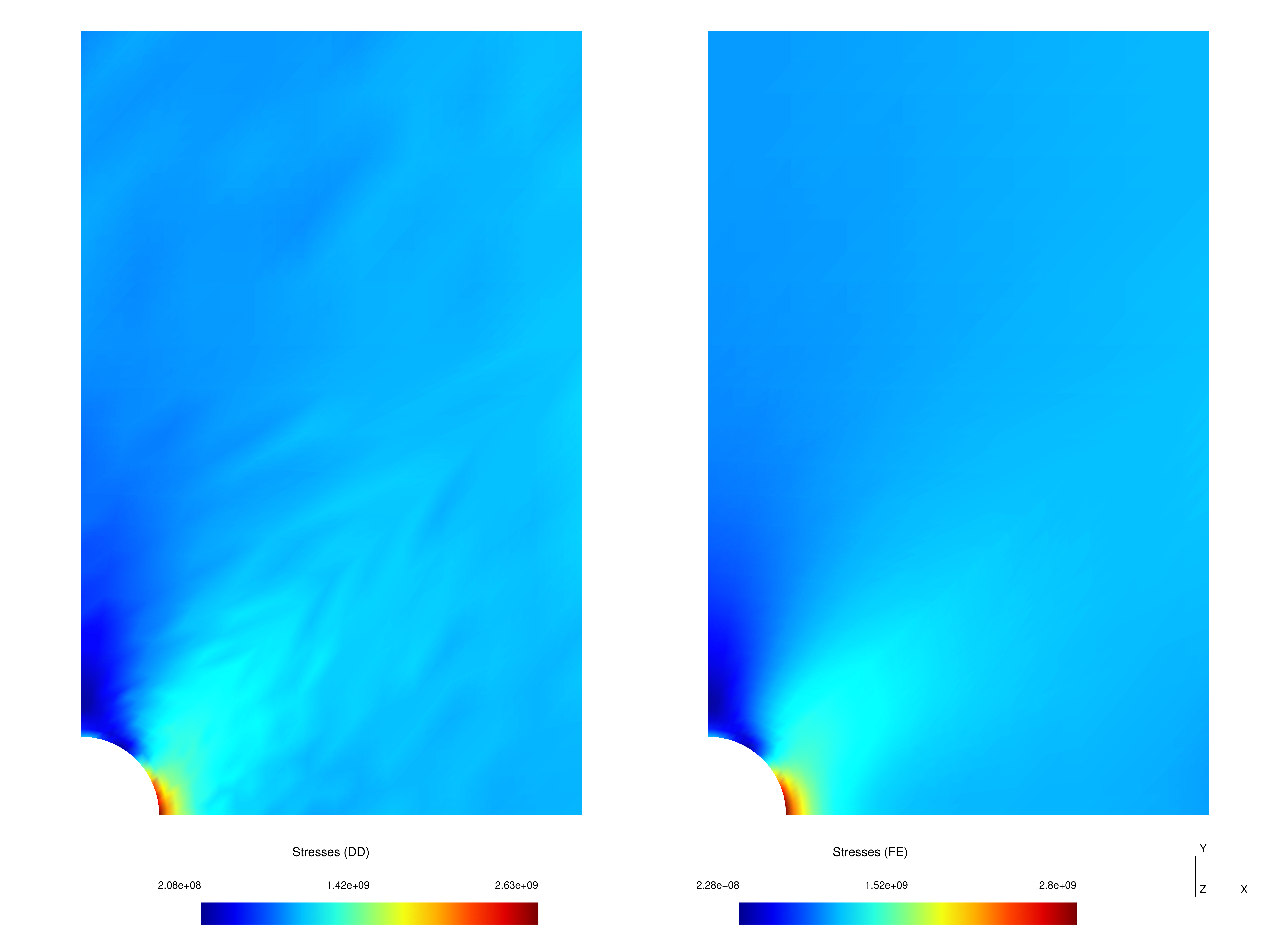}
\caption{Comparison of stress fields obtained by DDCM and FEM on the case of a plate with a hole (DDCM results obtained with DDI-DB2)}
\label{fig:plate-DDvsFE}       
\end{figure}

\begin{figure}[hbt]
\centering
\includegraphics[width=\linewidth]{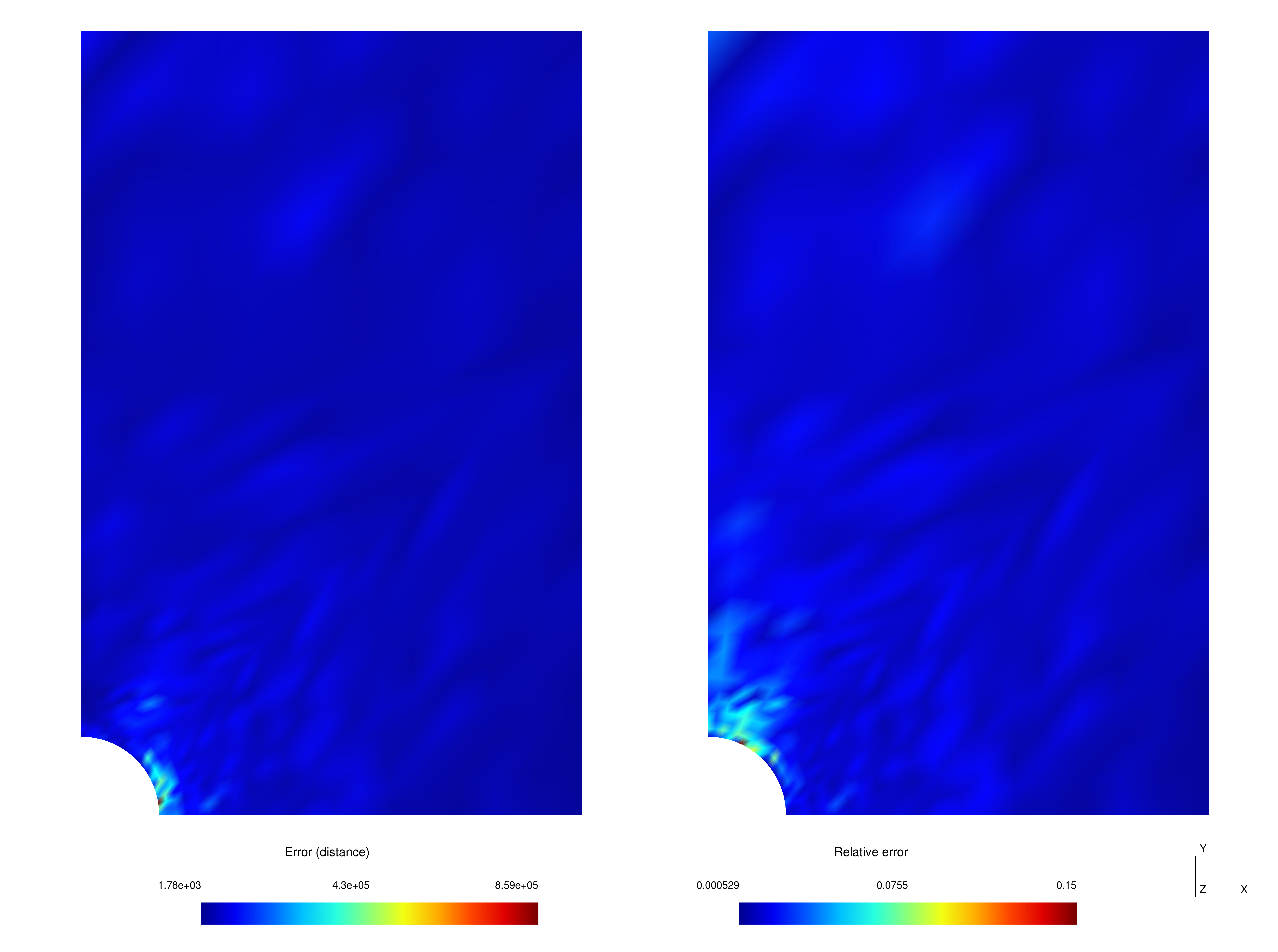}
\caption{Distance between DDCM and FEM solutions on the case of a plate with a hole (DDCM results obtained with DDI-DB2). The relative error is with respect to the local energy density $\tens{\epsilon}\cdot\tens{\sigma}$.}
\label{fig:plate-DDvsFE-error}       
\end{figure}

Figure \ref{fig:plate-DDvsFE} shows a comparison of the stress field obtained by DDCM with respect to our reference solution, obtained by FEM. We observe an overall good agreement, with the DDCM solution correctly picking the stress concentration at the side of the hole. The distribution of absolute and relative errors (in the metric of $\mathbb{C}_e$) is shown in Fig.\ \ref{fig:plate-DDvsFE-error}. For the database illustrated here (DDI-DB2), the error on the stress concentration amplitude is of the order of 10\%. This error can be reduced by enriching the database and/or tuning the metric tensor $\mathbb{C}_e$. Note that with the values which were (arbitrarily) chosen here, the error on strain fields, with respect to the FE solution, is much lower (of the order of 1\%, and less in most of the plate, for the case illustrated in Fig.~\ref{fig:plate-DDvsFE}) than the error on stress fields.

\begin{figure}[hbt]
\centering
\includegraphics[width=\linewidth]{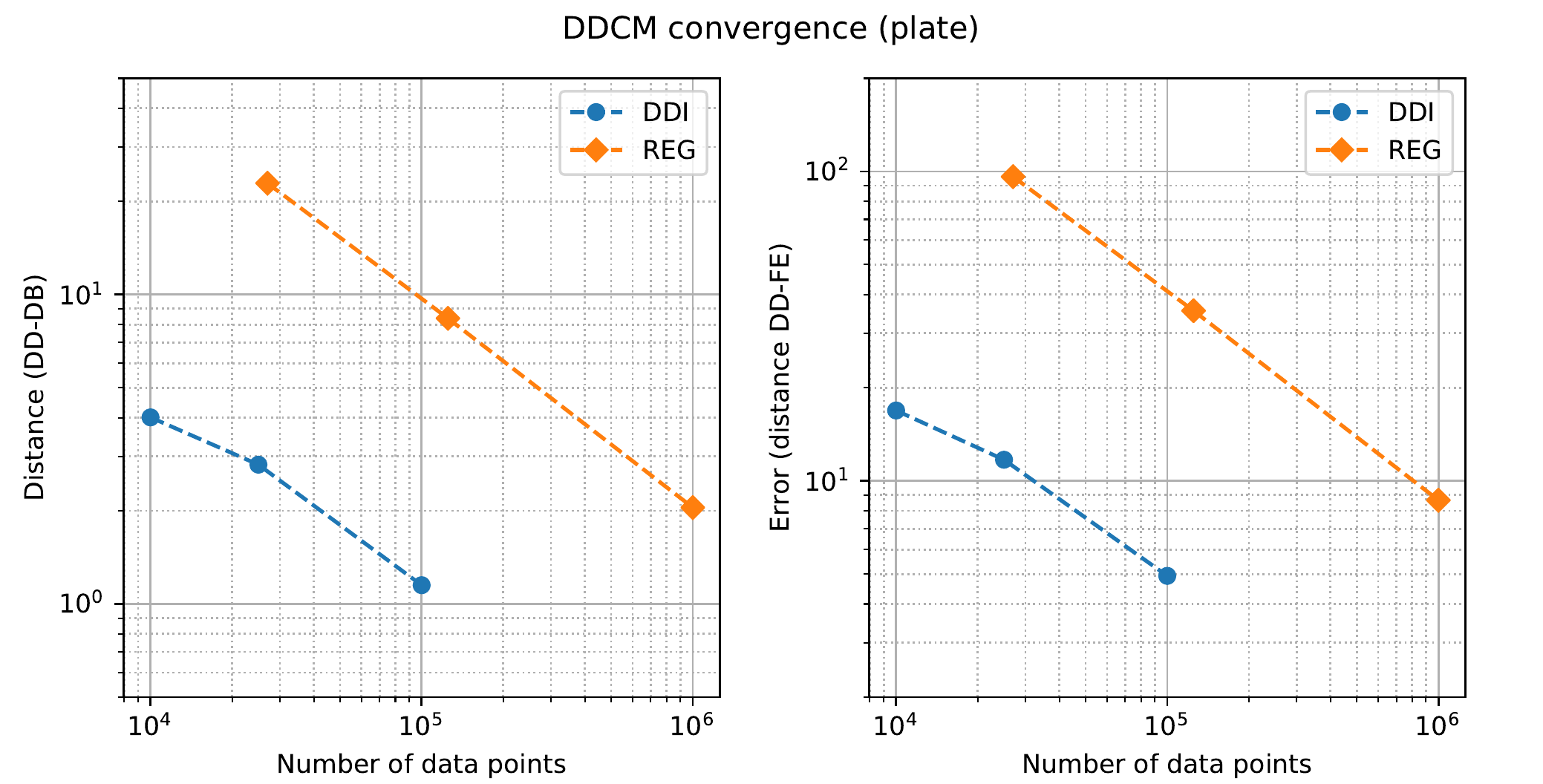}
\caption{Convergence of the DDCM method (plate example). Left plot shows convergence of the distance between the mechanical states and the database. Right plot shows the convergence of the error with respect to the reference FE solution.}
\label{fig:plate-DDCM-conv}       
\end{figure}

For a more quantitative comparison, we can look at distances between the various solution sets, in the metric of $\mathbb{C}_e$. Figure \ref{fig:plate-DDvsFE-error} shows the distribution of the local distance between DDCM and FEM solutions, while Table \ref{tab:distances-plate} lists global measures (i.e.\ integrated on the whole domain). From the first row in Table \ref{tab:distances-plate}, we see that DDI-generated databases are providing a better representation of, i.e.\ are closer to, the FE reference solution at a given database size. Note that the mechanical boundary-value problem considered here is different from those used to identify the databases.
From the third row in Table \ref{tab:distances-plate}, we also see that a DDI-generated database of 25,000 points (DDI-DB2) allows to obtain a DDCM solution of similar quality as a 1,000,000 point database using uniform sampling (REG-DB3), providing a strong argument in favour of using the former. 
For reference, recall that the total elastic strain energy computed in the reference FE solution is 737 J.
Figure \ref{fig:plate-DDCM-conv} shows the convergence of the minimal distance attained by the DDCM algorithm, as a function of the database size, and of the error with respect to the FE reference solution. Both curves clearly show that DDI databases significantly outperform REG ones, not only at equivalent size but overall. Both sets of databases appear to have similar asymptotic convergence rates.

%
\begin{table*}[htb]
\caption{Distance between solutions and material database (distances in energy [J]) -- plate example}
\label{tab:distances-plate}       
\begin{tabular}{lcccccc}
\hline\noalign{\smallskip}
Distance & DDI-DB1 & DDI-DB2 & DDI-DB3 & REG-DB1 & REG-DB2 & REG-DB3  \\
\noalign{\smallskip}\hline\noalign{\smallskip}
FEM-DB & 1.30 & 0.70 & 0.28 & 5.86 & 1.87 & 0.45 \\
DDCM-DB & 4.01 & 2.82 & 1.15 & 22.91 & 8.38 & 2.05 \\
DDCM-FEM & 16.88 & 11.70 & 4.93 & 96.20 & 35.47 & 8.66 \\
\noalign{\smallskip}\hline
\end{tabular}
\end{table*}

\subsubsection{L-beam}

As a second example, we consider a L-shaped beam (plane strain assumption) with the following geometrical characteristics: total width $W=0.6\,H$, vertical branch of width $w=0.2\,H$, horizontal branch of height $h=0.3\,H$, total height $H=1$m. The fillet has a curvature radius $r=0.02\,H$. A hole of radius $R=0.075H$ is positioned with its center at $(0.2\,H,0.5\,H)$. The base is fixed so as to prevent rigid body modes, and a horizontal displacement of $0.002\,H$ is imposed at the top. A  reference solution is obtained by using a classical FE simulation, using the same elastic properties as those used to generate the input images for the DDI method above (i.e.\ $E=217.5$ GPa, $\nu=0.3$, plane strain), on a mesh of 847 quadratic triangular elements (1823 nodes).
The total elastic strain energy associated to the reference FE solution is 1510 J. This number will constitute a useful reference when discussing the distances to and from material databases in the following.

This problem was simulated using the DDCM method described in section \ref{sec:algo}, using the different material databases identified by DDI (see above), and the same mesh as for the FE reference solution. As in the first example, we also used the three purely synthetic databases, labelled REG-DB1, REG-DB2, and REG-DB3, for comparison with DDI. The geometry and loading of this test case differ more significantly than in the previous one from those of the sample used for identification, which should provide a more straining test for the DDI databases.

\begin{figure}[hbt]
\centering
\includegraphics[width=\linewidth]{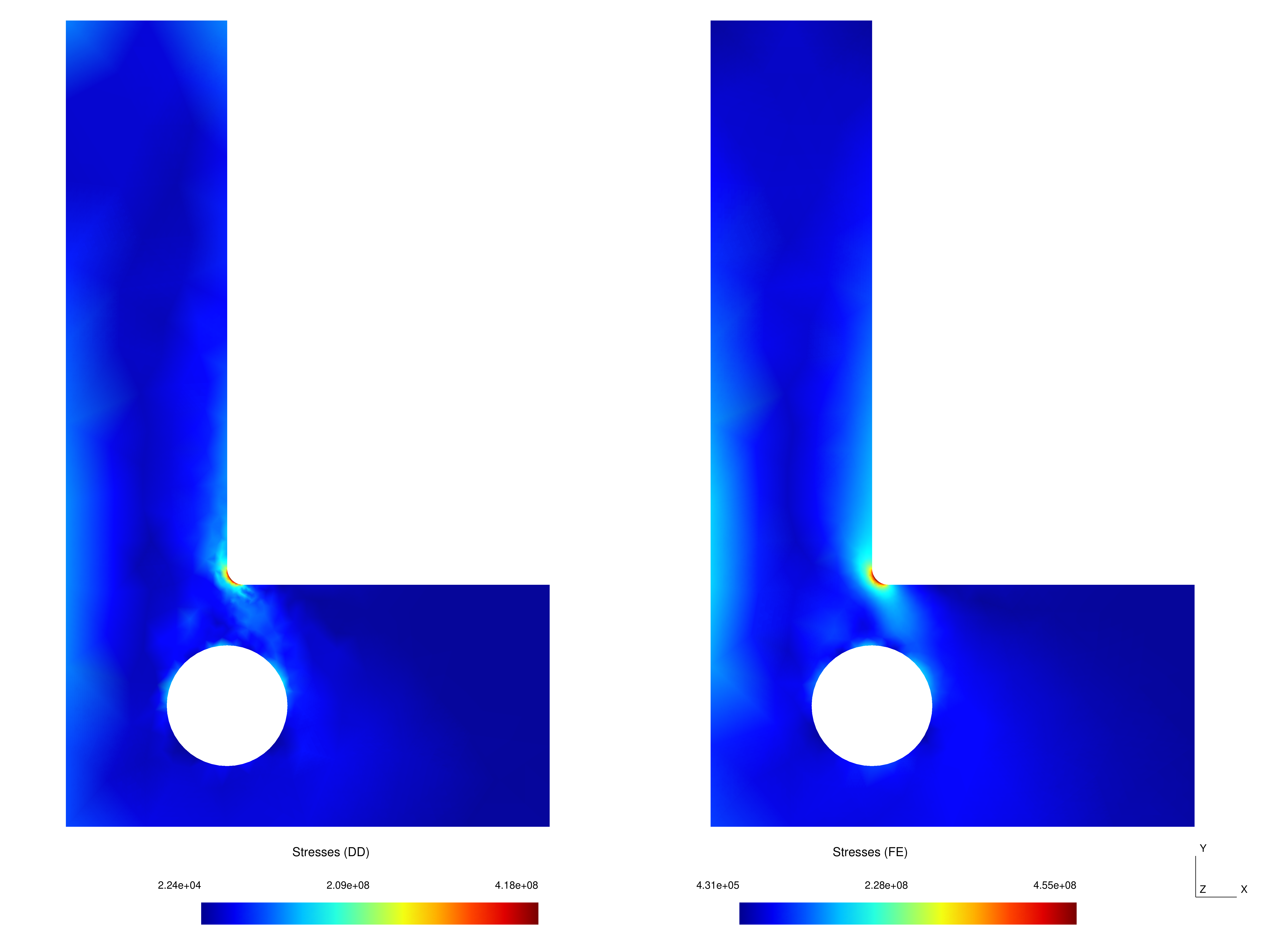}
\caption{Comparison of stress fields obtained by DDCM and FEM on the case of a L-beam (DDCM results obtained with DDI-DB2)}
\label{fig:beam-DDvsFE}       
\end{figure}

\begin{figure}[bht]
\centering
\includegraphics[width=\linewidth]{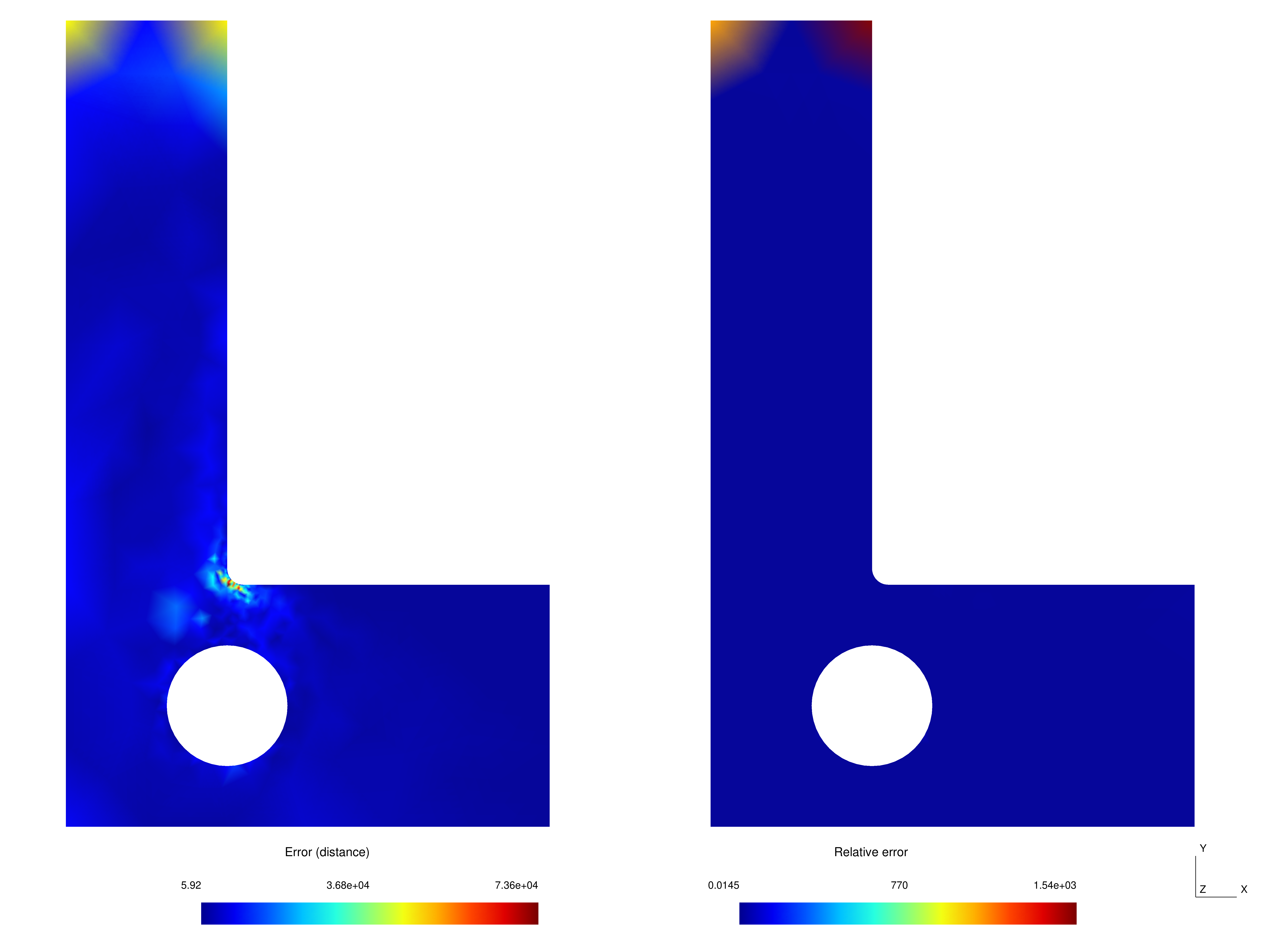}
\caption{Distance between DDCM and FEM solutions on the case of a L-beam (DDCM results obtained with DDI-DB2). The relative error is with respect to the local energy density $\tens{\epsilon}\cdot\tens{\sigma}$.}
\label{fig:beam-DDvsFE-error}       
\end{figure}

Figure \ref{fig:beam-DDvsFE} shows a comparison of the stress field obtained by DDCM with respect to our reference solution, obtained by FEM. We observe an overall good agreement, with the DDCM solution correctly picking the stress concentration at the fillet. Figure \ref{fig:beam-DDvsFE-error} shows the distribution of absolute and relative errors (in the metric of $\mathbb{C}_e$). The maximal absolute error is located at the fillet, but the maximal relative error actually occurs at the top corners, where the loading (imposed displacement) is applied.

\begin{figure}[hbt]
\centering
\includegraphics[width=\linewidth]{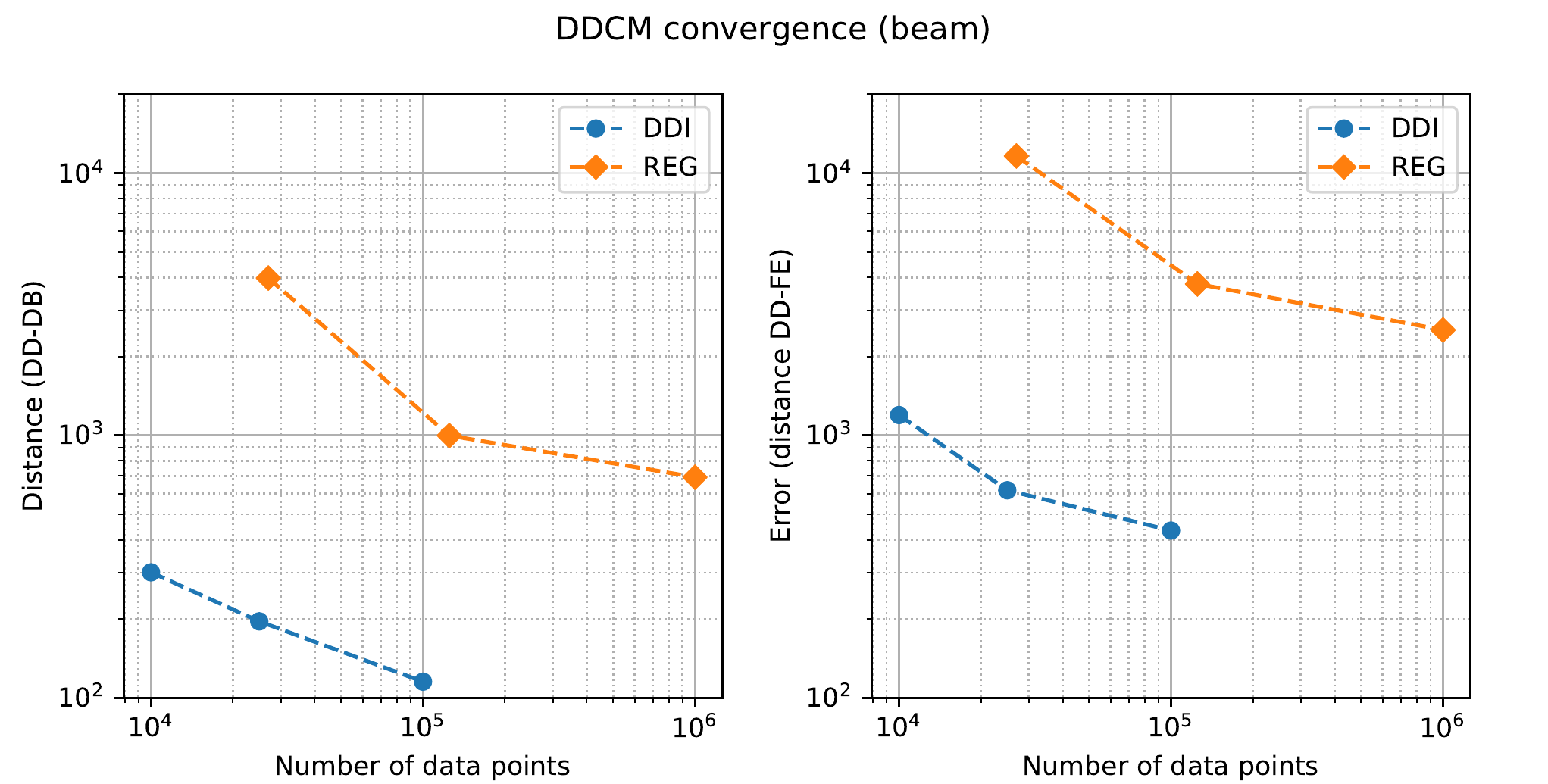}
\caption{Convergence of the DDCM method (beam example). Left plot shows convergence of the distance between the mechanical states and the database. Right plot shows the convergence of the error with respect to the reference FE solution.}
\label{fig:beam-DDCM-conv}       
\end{figure}

For a more quantitative comparison, we can look again at distances between the various solution sets, in the metric of $\mathbb{C}_e$. Figure \ref{fig:beam-DDvsFE-error} shows the distribution of the local distance between DDCM and FEM solutions, while Table \ref{tab:distances-beam} lists global measures (i.e.\ integrated on the whole domain). From the first row in Table \ref{tab:distances-beam}, we see that, like in previous example, DDI-generated databases are providing a better representation of, i.e.\ are closer to, the FE reference solution at a given database size.
From the third row in Table \ref{tab:distances-beam}, we also see that a DDI-generated database of 25,000 points (DDI-DB2) allows to obtain a DDCM solution of better quality than a 1,000,000 point database using uniform sampling (REG-DB3), providing once more a strong argument in favour of using the former. Figure \ref{fig:beam-DDCM-conv} shows the convergence of the minimal distance attained by the DDCM algorithm, as a function of the database size, and of the error with respect to the FE reference solution. Both curves clearly show that DDI databases significantly outperform REG ones, even with a limited number of data points.

%
\begin{table*}[htb]
\caption{Distance between solutions and material database (distances in energy [J]) -- beam example}
\label{tab:distances-beam}       
\begin{tabular}{lcccccc}
\hline\noalign{\smallskip}
Distance & DDI-DB1 & DDI-DB2 & DDI-DB3 & REG-DB1 & REG-DB2 & REG-DB3  \\
\noalign{\smallskip}\hline\noalign{\smallskip}
FEM-DB & 329.68 & 174.18 & 91.18 & 3434.35 & 1014.91 & 303.48 \\
DDCM-DB & 300.72 & 195.52 & 115.15 & 3976.76 & 998.25 & 693.04 \\
DDCM-FEM & 1195.73 & 617.84 & 433.73 & 11616.54 & 3779.12 & 2522.43 \\
\noalign{\smallskip}\hline
\end{tabular}
\end{table*}

\subsubsection{Material symmetries}


If it is \emph{a priori} known that the material behavior is isotropic, associated symmetries can be exploited and the intrinsic dimension of the material database can be reduced to that of invariants (3 for strains and 3 for stresses in 3D) \cite{KirchdoerferOrtiz2016}. As a consequence, the distance used in algorithm \eqref{eq:DD_solver} can be modified to
\begin{equation}
  \hat{d}(\vec{z},\vec{y}) = \min_{\{\tens{R}_e\}} \Big( \sum_{e=1}^{M} 
                                                         w_e | \vec{z}_e - \hat{\vec{y}}_e |_e^2 \Big)^{\frac{1}{2}}
\end{equation}
where $\hat{\vec{y}}_e = \{ \hat{\tens{\epsilon}}^*_e, \hat{\tens{\sigma}}^*_e \} 
                    = \{ \tens{R}_e^T\tens{\epsilon}^*_e\tens{R}_e, \tens{R}_e^T\tens{\sigma}^*_e\tens{R}_e \}$,
with $\tens{R}_e \in SO(3)$ a rotation tensor associated to material point $e$. When $\tens{\epsilon}_e$ and $\tens{\sigma}_e$ share the same eigenvectors, the rotation tensor can easily be derived analytically, but this is generally not the case, since they are updated independently from \eqref{eq:DDCM-eps} and \eqref{eq:DDCM-sigma}. The minimization can then be performed numerically (this operation is purely local for each material point), using a parametrization of the rotation tensor. In the 2D case considered in the above examples, such parametrization reduces to a single angle, and the minimization problem can be solved analytically.

Results obtained using this isotropic distance are shown in Table \ref{tab:distances-isotropic-plate}, for the case of the plate and DDI databases, and Table \ref{tab:distances-isotropic-beam}, for the case of the L-beam and DDI databases. When compared to values obtained with the standard distance (Tab. \ref{tab:distances-plate} and \ref{tab:distances-beam}), we see a reduction of the minimum distance, but the error with respect to the reference FE solution does not systematically decrease (e.g. in the case of the plate with DDI-DB1). Moreover, this improvement in the minimum distance typically comes at the cost of more iterations in the DDCM algorithm. The computational cost {\slshape vs.}\ precision trade-off is thus not always beneficial. Here, this approach appears more relevant for the beam case, where the  loading differs more significantly from that of the identification sample than in the case of the plate. Accounting for full $SO(3)$ orbits of material points in the database then enrich the latter, leading to improved results of the DDCM algorithm. Yet, this requires the \textsl{a priori} assumption of isotropy of the material.

\begin{table}[htb]
\caption{Minimal isotropic distance reached by DDCM and standard distance between DDCM and FEM solutions (distances in energy [J]) -- plate example}
\label{tab:distances-isotropic-plate}       
\begin{tabular}{lccc}
\hline\noalign{\smallskip}
Distance & DDI-DB1 & DDI-DB2 & DDI-DB3 \\ 
\noalign{\smallskip}\hline\noalign{\smallskip}
DDCM-DB & 2.53 & 1.45 & 0.44 \\ 
DDCM-FEM & 33.13 & 10.43 & 3.10 \\ 
\noalign{\smallskip}\hline
\end{tabular}
\end{table}

\begin{table*}[htb]
\caption{Minimal isotropic distance reached by DDCM and standard distance between DDCM and FEM solutions (distances in energy [J]) -- beam example}
\label{tab:distances-isotropic-beam}       
\begin{tabular}{lcccccc}
\hline\noalign{\smallskip}
Distance & DDI-DB1 & DDI-DB2 & DDI-DB3 \\ 
\noalign{\smallskip}\hline\noalign{\smallskip}
DDCM-DB & 89.26 & 45.26 & 18.10 \\ 
DDCM-FEM & 420.71 & 200.24 & 68.91 \\ 
\noalign{\smallskip}\hline
\end{tabular}
\end{table*}

\section{Discussion and perspectives}
\label{concl}

In this paper, we have shown how the DDI algorithm and DDCM solver naturally combine to constitute a computational design toolbox eliminating epistemic uncertainty linked to traditional constitutive models in mechanics. Not only does DDI provide a way to infer stresses directly from experimental field measures, where classical model updating techniques must postulate a specific constitutive model, it also generates databases of material states respecting some importance sampling. This latter feature proved to yield a greatly improved efficiency for DDCM in the examples considered.

DDCM also proved its robustness in the above examples, by its capacity to perform simulations on various geometries and loadings using a given database. Qualitatively, the method requires only limited data to predict the major features of strain and stress fields (e.g. location of maximal values). Quantitatively, the precision obtained depends directly on the quality of the database (number and sampling of data points), but one should note that the DDCM comes with its own error estimate. Indeed, the minimal distance which can be attained between mechanical states (i.e.\ those verifying compatibility and balance constraints) and material states (i.e.\ those in the database) provides a direct indicator of the adequacy of the given database for the considered mechanical problem. If this distance cannot be sufficiently decreased by the algorithm, it can be considered as an indication that more points are needed in the database.

Linear systems occurring in DDCM algorithm correspond to standard elasticity problems with pre-strain/pre-stress loading. This opens a perspective to implementing this step through third-party FE software computing engines. These softwares would need to allow for individual specification of pre-strain / pre-stress at each integration point. If this feature seems available in the open-source software Code\_Aster \cite{Code_Aster} for example, it may require going through user-defined elements in most commercial softwares.

In practice, most of the computational cost resides in database searches. For linear system solutions, a single factorization is typically necessary, even in the case of a multi-step simulation with non-linear material behavior, unless one wants to adapt the metric at each step, which is of course possible. Thus, although the above examples were treated using a basic search algorithm, computational efficiency could significantly be improved by using data structures and search algorithms adapted to manipulating and navigating very large data sets (e.g.\ \cite{MujaLowe2014}).

It is also important to note that the whole DDI+DDCM framework remains valid, without any modifications, in the presence of material non-linearities. Indeed, those will be entirely contained within the database, and the linearity of the systems of equations solved in DDI and DDCM is unaffected by material non-linearities. Those will thus have no effect on computational performance either. For material behaviors presenting irreversibilities and/or history dependence, first results have been established, both in DDI \cite{Leygue_etal2019} and DDCM \cite{Eggersman_etal2019}.

Finally, the Data-Driven paradigm, and in particular the DDI+DDCM framework, opens promising perspectives in terms of public repositories of material databases. Such web-based platforms would allow and encourage capitatlization and sharing of large amounts of material data, experimental and/or synthetic (e.g.\ multiscale computations), which are typically thrown away after constitutive model identification in the current dominant model-based paradigm. Of course, this would probably require a preliminary stage of standardization of database formats (including meta-data) and curation methods by the community.


\bibliographystyle{spmpsci}      
\bibliography{biblio}   

%
%

\end{document}